\documentclass[pre,twocolumn,floatfix,superscriptaddress,10pt]{revtex4}
\usepackage{graphics}
\usepackage{amssymb}
\usepackage{epsfig}
\usepackage{subfigure}
\usepackage{color}
\usepackage{vmargin}
\setmarginsrb{1.5cm}{1cm}{1.5cm}{2cm}{0mm}{10mm}{0mm}{10mm}

\usepackage{fancyhdr}
\fancypagestyle{title}{
  \setlength{\headheight}{0pt}%
  \fancyhf{}
  \fancyfoot[L]{\scriptsize 1050-2947/2012/85(4)/045803(3)}
  \fancyfoot[C]{\scriptsize 045803-\thepage}
  \fancyfoot[R]{\scriptsize \copyright2012 American Physical Society}
  \fancyhead[C]{\scriptsize PHYSICAL REVIEW A {\bf 85}, 045803 (2012)}
}

\begin{document}
\title{Exploiting bifurcations in waveguide arrays for light detectors}
\author{N.\ Karjanto}
\email{natanael@skku.edu}
\affiliation{Department of Mathematics, University College\\ Sungkyunkwan University, Natural Science Campus, South Korea}
\author{H.\ Susanto}
\email{hadi.susanto@nottingham.ac.uk}
\affiliation{School of Mathematical Sciences, University of Nottingham\\ University Park, Nottingham, NG7 2RD, United Kingdom\\
{\normalfont \footnotesize (Received 20 September 2011; published 16 April 2012)}
}

\begin{abstract}
An array of a finite number waveguides, driven laterally by injecting light at the outer waveguides, is considered. The array is modeled by a discrete nonlinear Schr\"{o}dinger equation. It has been shown [{\color{blue} Phys. Rev. Lett. {\bf 94}, 243902 (2005)}] that, when the injected light is in the proximity of a bifurcation point, such a system can be sensitive to small disturbances, making it possible to act as a light detector. Here, the optimum intensity of the injected light is discussed, and an analytical approximation is presented. \vspace{0.2cm}\\
{\footnotesize DOI: {\color{blue} 10.1103/PhysRevA.85.045803} \hfill PACS number(s): 42.65.Tg, 05.45.Yv, 42.82.Et}
\end{abstract}
\keywords{waveguide arrays; light detectors; discrete nonlinear Schr\"{o}dinger equation.}

\maketitle
\thispagestyle{title}

The idea of discretizing the continuous behaviors of electromagnetic fields emerged rather gradually in the area of optics. One possible scenario is achieved by evolving the fields through waveguide arrays. One of the first seminal papers in the study of the discrete behaviors of lights in coupled waveguides is due to 
Jones~\cite{jone65}. It was shown that the waveguide transversal field may experience diffraction different from that occurring in continuous systems where the energy splits into two main branches with a set of secondary peaks between them. The evolution was shown to be described by a Bessel function of the first kind~\cite{jone65}. An experimental observation of this interesting result was reported in Ref.~\cite{some73} in a system of GaAs waveguide arrays.

In addition to discrete diffractions, nonlinear waveguide arrays may also support self-localization in the form of optical discrete solitons, which was predicted by Christodoulides and Joseph~\cite{chri88}. The prediction was confirmed in Ref.~\cite{eise98} in experiments that were carried out in a highly nonlinear AlGaAs waveguide array. In subsequent papers, it was shown that discrete solitons can be effectively routed and can be blocked using essentially soliton collisions~\cite{chri01,chri01_2,euge01}. Hence, among others, waveguide arrays provide a rich environment for all-optical data-processing applications, such as logic functions and time gating. The reader is referred to recent papers~\cite{lede08,kevred09} on the topic.

Recently, another application of waveguide arrays was proposed in Ref.~\cite{khomer05}. A system composed of a finite number of waveguide arrays driven by a linear waveguide at the boundaries was considered. The light localization at the boundaries is similar to that in discrete surface solitons~\cite{sunt06}. It was proposed that such an arrangement can be sensitive to small disturbances when the amplitude of the lateral driving light is such that the system is close to a bifurcation point. In discussing the finding analytically, the authors of Ref.~\cite{khomer05} consider the continuous limit of the governing equation, i.e., a discrete nonlinear Schr\"{o}dinger equation, which is accurate when the array has enough waveguides and strong coupling. Even though the same property was also observed in a fully discrete case, i.e., a few waveguides or small coupling, in Ref.~\cite{khomer05}, it is stated that they do not have, in that case, an analytical description. Here, we provide an analysis where the system is rather genuinely discrete. The method is similar to that in Ref.~\cite{susa08} used to describe the so-called supratransmission~\cite{leon03,Khomeriki} or self-induced transparency~\cite{mani06} phenomenon in discrete systems.

In the following, we will first describe the governing equation of the problem. We then present the theoretical analysis explaining the reported property. Finally, we give conclusions and remarks for our discussion.

We consider a system of waveguide arrays that is described by the following dimensionless discrete Schr\"odinger equation with Kerr nonlinearity~\cite{khomer05}:
\begin{equation}
\displaystyle
\left(i\partial_z+i\gamma-2Q-V\right){\psi}_n+Q\left(\psi_{n-1}+\psi_{n+1}\right)+|\psi_n|^2\psi_n=0,
\label{eq1}
\end{equation}
where $n = 2$--$4$, laterally driven by linear waveguides,
\begin{equation}
\displaystyle
i\left(\partial_z+\gamma_0\right){\psi}_1=Q_0\psi_{2},\quad i\left(\partial_z+\gamma_0\right){\psi}_5=Q_0\psi_{4}.
\label{lat1}
\end{equation}
The dependent variable $\psi_n$ denotes the complex electric-field envelope, $Q$ and $Q_0$ are the linear coupling between the waveguides, and $V$ is the on-site potential of the inner waveguides. The index difference $V$ between the inner and the outer waveguides is introduced such that the guided modes in the two lateral waveguides would not linearly propagate in the array \cite{khomer05}. The parameters $\gamma$ and $\gamma_0$ are the attenuation parameters representing field losses in the waveguides. 

\begin{figure}[tbhp]
  \begin{center}
    \includegraphics[width=0.45\textwidth,angle=0]{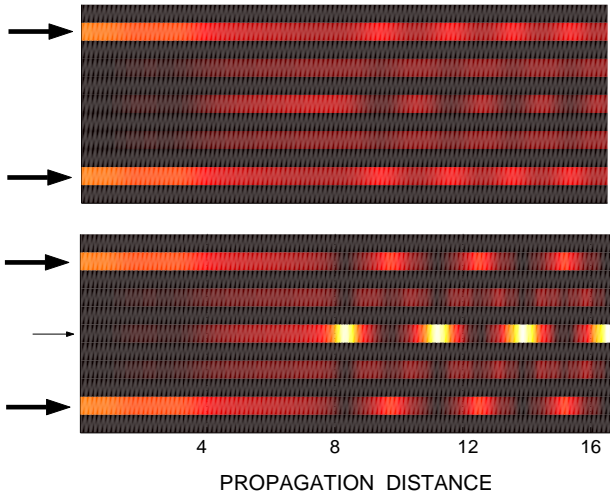}
  \end{center}
  \caption{(Color online) Flux intensity along the waveguides driven at the threshold of bistability (large arrows) obtained from Eq.~(\ref{dyn}). Here, $Q = Q_0 = 4$, $\gamma = \gamma_0 = 0.001$, $V = 0.5$, and $A = 3.0946$. In the lower panel, it is shown that a perturbation of amplitude $0.001$ (small arrow) is enough to switch and to amplify the field of the middle waveguide. All given quantities are dimensionless. The figure and the information of the parameter values are taken from Ref.~\cite{khomer05} with permission from the authors.}
  \label{dyn}
\end{figure}

In Fig.~\ref{dyn}, the dynamics of the wave field for initial conditions $\psi_1(0) = \psi_5(0) = A$ and $\psi_n(0) = 0, \, n \neq 1,5$ is presented with the driving amplitude $A$ close to a critical value. When there is no disturbance, one could observe that the fields are minimum in the middle waveguide. Yet, it is interesting to note that, when there is a small wave field in the middle waveguide, the dynamics is modified and the electromagnetic field of the middle waveguide is amplified as shown in the bottom panel of the same figure. This is the behavior that was proposed as an ultrasensitive light detector in Ref.~\cite{khomer05}.

To explain the phenomenon analytically, we consider a modified expression of the lateral waveguides (\ref{lat1}), i.e., we assume~\cite{khomer05}
\begin{equation}
\psi_1(z) = \psi_5(z)= A.
\end{equation}
Even though the intensity of the lights tunneling in the outer waveguides decays along $z$, the assumption can be expected to be valid if the coupling constant $Q_0$ and the attenuation $\gamma_0$ are small. To simplify the analysis, we also assume that $\gamma=\gamma_0=0$ \cite{khomer05} and $Q=Q_0$. The time-independent governing equation (\ref{eq1}) is then given by
\begin{equation}
\begin{array}{lll}
\displaystyle V \psi_2 - Q(A + \psi_3 - 2 \psi_2)- \psi_2^3 &=& 0,\\
\displaystyle V \psi_3 - Q(\psi_2 + \psi_4 - 2\psi_3) - \psi_3^3 &=& 0,\\
\displaystyle V \psi_4 - Q(\psi_3 + A - 2\psi_4) - \psi_4^3 &=& 0.
\end{array}
\label{gov2}
\end{equation}

As discussed in Ref.~\cite{khomer05}, using a \emph{continuous approximation} of Eq.~(\ref{gov2}) to act as a light detector of a small disturbance, the driving amplitude $A$ must be in the vicinity of a critical driving $A_{\text{crit}}$ at which the time-independent solution of Eq.~(\ref{gov2}), corresponding to the profile in the top panel of Fig.~\ref{dyn}, disappears in a saddle-node bifurcation. In the following, we will show that one can also use the the \emph{discrete approximation} (\ref{gov2}) to explain the mechanism of the phenomenon as well as approximate the critical driving amplitude.

To obtain an analytical approximation for $A_{\text{crit}}$ using Eq.~(\ref{gov2}), we scale the equations such that they become
\begin{equation}
\begin{array}{lll}
\displaystyle \psi_2 - \epsilon(A + \psi_3 - 2\psi_2 + { \delta}\psi_2^3) &=& 0,\\
\displaystyle \psi_3 - \epsilon(\psi_2 + \psi_4 - 2\psi_3 + {\delta}\psi_3^3) &=& 0,\\
\displaystyle \psi_4 - \epsilon(\psi_3 + A - 2\psi_4 + {\delta}\psi_4^3) &=& 0,
\end{array}
\label{gov3}
\end{equation}
where $\epsilon = Q/V$ and $\delta = 1/Q$. Our analysis is based on a formal perturbation expansion in $\epsilon$ by assuming that $0 < \epsilon \ll 1$ and $\delta \sim O(1)$.

Next, we write $A$, $\psi_{2}$, $\psi_{3}$, and $\psi_{4}$ in the following perturbation expansion:
\begin{eqnarray}
  A &=& A_{0} \epsilon^{-3/2} + A_{1} \epsilon^{-1/2} + O\left(\epsilon^{1/2}\right), \\
  \psi_{2} = \psi_{4} &=& K_{1} \epsilon^{-1/2} + K_{2} \epsilon^{1/2} + O\left(\epsilon^{3/2}\right), \\
  \psi_{3} &=& L_{1} \epsilon^{1/2} + L_{2} \epsilon^{3/2} + O\left(\epsilon^{5/2}\right).
\end{eqnarray}
Substituting them into~Eq.\ (\ref{gov3}) and collecting the terms accordingly, we obtain series expansions which are identical for the cases of $n = 2$ and $n = 4$. For the equations of $O\left(\epsilon^{-1/2}\right)$ and $O\left(\epsilon^{1/2}\right)$, the relationships are held as follows:
\begin{eqnarray}
  \delta \, K_{1}^{3} - K_{1} + A_{0} &=& 0, \label{cubic} \\
  3 \, \delta \, K_{1}^{2} K_{2} - 2K_{1} - K_{2} + A_{1} &=& 0. \label{setengah}
\end{eqnarray}

Equation (\ref{cubic}) is a cubic equation in $K_1$ with the number of roots depending on $A_0$. It can be easily calculated that the critical $A_0$ at which there is a transition from the cubic equation having three roots to one root is
\begin{equation}
A_{0,\text{crit}} = 2/\sqrt{27\delta}.
\end{equation}
At this value of $A_0$, the critical root of the cubic equation~(\ref{cubic}) is
\begin{equation}
K_{1,\text{crit}} = 1/\sqrt{3\delta}.
\end{equation}
Substituting those critical values for $A_0$ and $K_1$ into Eq.~(\ref{setengah}), we obtain the critical value for $A_{1}$,
\begin{equation}
A_{1,\text{crit}} = 2/\sqrt{3\delta}.
\end{equation}
One can continue the calculations further to obtain higher-order corrections of the critical driving amplitude $A_{\text{crit}}$; we leave the details to the interested reader. For the leading order, we obtain
\begin{equation}
A_{\text{crit}}=
{\frac {2\sqrt {3}}{9\sqrt {\delta}{\epsilon}^{3/2}}}+{\frac {2\sqrt {3}}{
3\sqrt {\delta\epsilon}}}-{\frac {\sqrt {3\epsilon}}{3\sqrt {\delta}}}+
{\frac {5\sqrt {3}{\epsilon}^{3/2}}{9\sqrt {\delta}}}
+ \cdot\cdot\cdot \, .
\label{appr}
\end{equation}
This is our main result.

\begin{figure}[tbhp]
  \begin{center}
    \includegraphics[width=0.5\textwidth]{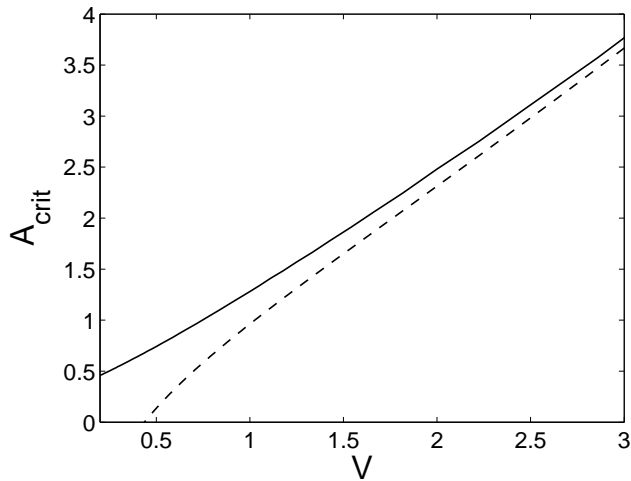}
  \end{center}
  \caption{Comparison between the numerically obtained critical value $A_{\text{crit}}$ from Eq.~(\ref{cubic}) (solid line) and the approximation (\ref{appr}) (dashed line) as a function of $V$ for $Q=1$. }
  \label{comp}
\end{figure}

In addition to the existence analysis above, one may also calculate the stability of the calculated solutions perturbatively. Let $\phi_n$, $n =1$--$3$, be a solution of Eq.~(\ref{gov2}). The linear stability of $\phi_n$ can be obtained by substituting the ansatz $\psi_n = \phi_n + \epsilon [v_n e^{i\lambda z} + \overline{w_n} e^{-i\overline{\lambda}z}]$ with $\lambda \in \mathbb{C}$ and $(v_n, w_n) \in \mathbb{C}^2$ into Eq.~(\ref{eq1}) (with $\gamma=0$). Linearizing the equation to $O(\epsilon)$, one will obtain the eigenvalue problem,
\begin{equation}
		\lambda\epsilon\delta\left(\begin{array}{cc}v_n\\ w_n\end{array}\right) =
		\epsilon\left(\begin{array}{cc}v_{n-1}+v_{n+1}\\ -w_{n-1}-w_{n+1}\end{array}\right) +
		\mathcal{L}_n\left(\begin{array}{cc}v_{n}\\ w_{n}\end{array}\right),
\label{eig}
\end{equation}
with 
\[
\left(\begin{array}{c} v_{0,4}\\w_{0,4}\end{array} \right)=\left(\begin{array}{c} 0\\0\end{array} \right) 
\]
and
\begin{eqnarray}
\mathcal{L}_n=
\left(
\begin{array}{cc}
-1-2\epsilon+2\epsilon\delta|\phi_n|^2 & \epsilon \delta \phi_n^2 \\
-\epsilon\delta\overline{\phi_n}^2 & 1+2\epsilon-2\epsilon\delta|\phi_n|^2
\end{array} \right).\nonumber
\end{eqnarray}
$\phi_n$ is linearly stable if the imaginary part of $\lambda$ is zero, i.e., Im$(\lambda)=0$.

Using the same perturbation technique as above (see also Ref.~\cite{susa08}), for the leading order, the eigenvalue problem (\ref{eig}) can be approximated by
\[
\lambda\epsilon\delta\left(\begin{array}{cc}v_{\{1,3\}}\\ w_{\{1,3\}}\end{array}\right) = \mathcal{L}_{\{1,3\}}\left(\begin{array}{cc}v_{\{1,3\}}\\ w_{\{1,3\}}\end{array}\right),
\]
which gives the following approximate eigenvalue:
\begin{equation}
\left(\epsilon \delta \lambda \right)^2 = 1 - 4 \, \delta \, K_1^2 + 3 \, \delta^2 \, K_1^4 + O(\epsilon).
\label{lineig}
\end{equation}
It can be easily calculated that, at the saddle-node bifurcation, i.e., $K_1 = K_{1,\text{crit}}$, $\lambda = 0$. For $K_1 < K_{1,\text{crit}}$, $\lambda^2 > 0$ and vice versa. Hence, the turning point is indeed a standard saddle-node bifurcation where a stable solution collides with an unstable one. We also obtain a similar stability result as in Ref.~\cite{khomer05} that the solution shown in the upper plot in Fig.~\ref{dyn}, which corresponds to $K_1 < K_{1,\text{crit}}$, is stable.

We have solved the discrete equation~(\ref{gov2}) numerically. In Fig.~\ref{comp}, we compare the numerical result of the critical driving amplitude and our approximation~(\ref{appr}) where one can note that the agreement is good when $V \gg 1$. We have also simulated the time-dependent equations~(\ref{eq1}) and~(\ref{lat1}) where we obtained that $A_{\text{crit}}$ from Eq.~(\ref{gov3}) above provides a lower bound to the critical amplitude of the original governing equation. Nonetheless, it is quite unfortunate that our prediction is not in good agreement with the data in Fig.~\ref{dyn}, which are taken from Ref.~\cite{khomer05}. We suspect that there might be typographical errors in the parameter values of Fig.~2 of Ref.~\cite{khomer05} as the authors referred the waveguide arrays in the figure to be genuinely discrete, yet the coupling constant $Q$ is relatively large and the on-site potential $V$ is small. We are also not able to reproduce the dynamics using the same parameter values. In addition to the existence analysis, we have calculated the stability of the solutions that collide and disappear in a saddle-node bifurcation numerically. We obtain an agreement with the analysis discussed above that the solution with a lower power is stable, and the critical eigenvalue $\lambda$ becomes zero at the turning point (not shown here).

To conclude, we have considered an array of a finite number of waveguides driven laterally by injecting light at the outer waveguides modeled by a discrete nonlinear Schr\"{o}dinger equation. Although, in this paper, we only consider five coupled equations, the method is applicable for larger systems. Note that the technique is adopted from Ref.~\cite{susa08}, which was used to study semi-infinite-coupled waveguide arrays. When the injected light is in the proximity of a bifurcation point, such a system can be sensitive to small disturbances. Here, we have discussed the optimum intensity of the injected light and have presented an analytical approximation of it.
\vfill
We acknowledge useful discussions with Ramaz Khomeriki and his courtesy for providing a copy of Fig.~\ref{dyn}.

\end{document}